\documentclass{iau} 
\usepackage{graphicx}

\title[PNe tracing the chemical the evolution of nearby dwarf galaxies] 
{What do planetary nebulae and H~{\sc ii} regions \\ reveal about the chemical evolution of \\ nearby dwarf galaxies?}
                
\author[Denise R. Gon\c calves]   
{Denise R. Gon\c calves $^{1,2}$
}

\affiliation{
$^1$Valongo Observatory, Federal University of Rio de Janeiro, \\ 
Ladeira Pedro Antonio 43, 20080-090 Rio de Janeiro, Brazil \\ 
$^2$East Asian ALMA Regional Center, National Astronomical Observatory of Japan \\ 2-21-1 Osawa, Mitaka, Tokyo, 181-8588, Japan \\
email: {\tt denise@astro.ufrj.br} 
}

\pubyear{2018}
\volume{344}  
\jname{Dwarf Galaxies: From the Deep Universe to the Present}
\editors{K. McQuinn, \& S. Stierwalt, eds.}
\begin{document}

\maketitle

\begin{abstract}
The Local Group contains a great number of dwarf irregulars and spheroidals, for
which the spectroscopy of individual stars can be obtained. Thus, the chemical evolution of these galaxies can be traced, with the only need of finding populations spanning a large age range and such that we can accurately derive the composition. Planetary nebulae (PNe) are old- and intermediate-age star remnants and their chemical abundances can be obtained up to 
3-4~Mpc. H~{\sc ii} regions, which are brighter and much easily detected, represent galaxies young content. PNe and H~{\sc ii} regions share similar spectroscopic features and are analysed in the same way. Both are among the best tracers of the chemical evolution allowing to draw the chemical time line of nearby galaxies. The focus in this review are the PN and H~{\sc ii} region populations as constraints to the chemical evolution models and the mass-metallicity relation of the local Universe.

\end{abstract}

\firstsection 
\section{Introduction}

Considering that this paper is aimed at reviewing the chemical evolution of the nearby dwarf galaxies, we concentrate on their most common types only: the dwarf irregulars (dIrr) and the dwarf spheroidals (dSph).  Dwarf ellipticals, dE (like NGC~205, \cite[Monaco et al. 2009]{Mon2009}; \cite[Gon\c calves et al. 2015]{gon2015}) will be included within the dSph, unless clearly stated otherwise. As it is well known, the most conspicuous properties that differentiate the dE, dSph and dIrr galaxies are: the H~{\sc i} amount ($\leq 10^8$~M$_{\odot}$, $\leq 10^5$~M$_{\odot}$ and $\leq 10^9$~M$_{\odot}$, respectively) and the total --stars plus gas-- mass ($\leq 10^9$~M$_{\odot}$, $\sim 10^7$~M$_{\odot}$ and $\leq 10^{10}$~M$_{\odot}$, respectively). The total to H~{\sc i} mass ratios thus imply that the dSph are dark-matter dominated while the dIrr are dominated by the atomic hydrogen. Being the star formation quiescent in the dSph and almost constant in the dIrr, the main stellar population in the former is of old- and intermediate-age, while the latter is predominantly populated by young stars. Though in different amounts, dE, dSph and dIrr show signatures of old stellar populations (\cite[Weisz et al. 2014]{wei2014}). It is also important to remember that whereas the dIrr tend to be isolated, the dSph galaxies in the Local Group are found to be satellites of the Milky Way and of the M31. 

Concerning the dwarf galaxy metallicities (Z), the above characteristics  also imply that for the star-forming dIrr the best Z indicator is the O/H (oxygen abundance) derived from the H~{\sc ii} regions. A few selected recent examples in the literature are: \cite[Berg et al. (2016)]{ber2016}; \cite[Pilyugin \& Grebel (2016)]{pigr2016}; \cite[Magrini et al. (2017)]{mag2017}; \cite[Flores-Dur\'an et al. (2017)]{fldu2017}; \cite[Kumari et al. (2018)]{kum2018}). Although some gas has been found in the central regions of dSph (e.g., in NGC~185, \cite[Gon\c calves et al. 2012]{gon2012}), these galaxies are dominated by the old stellar populations. As so, the metallicity of dwarf spheroidals is derived from the Ca~{\sc ii} triplet that gives the [Fe/H] abundance (\cite[Tolstoy et al. 2001]{tol2001}; \cite[de Boer et al. 2012]{deb2010}; \cite[Hansen et al. 2018]{han2018}). Therefore, any comparison between the metallicity of dIrr and dSph galaxies will need the conversion of one Z to another, the [Fe/O]. This abundance ratio is known to depend on the galaxy star formation history (SFH), as discussed by \cite[Richer \& McCall (1995)]{rimc1995}. For the sake of a \lq clean' Z comparison between these two types of galaxies, what about to adopt  metallicities from PNe that are present in star-forming (dIrr) and in quiescent (dSph) galaxies?

Our approach to study the metallicity evolution of the dwarf galaxies (mainly within the Local Group, LG) is based on  the adoption of the oxygen abundance of planetary nebulae as the main Z indicator. Our final goal is to derive the fundamental mass-metallicity relation (MZR) for these galaxies, by means of the their PN population. For that end, we first note that the chemical abundances of PNe and H~{\sc ii} regions are obtained in the same way, so providing results that can be robustly compared (Sections 2 and 3). Then, in Sect. 4, we review some of the most important results of PNe and H~{\sc ii} regions in terms of the metallicity gradients (or simply the Z distribution) in the nearby galaxies. Our own data, as well as a compilation from the literature are used to discuss a robust mass-metallicity relation, in Sections 5 and 6, which leads us to the remarkable suggestion that the MZR is actually universal in the local volume. This conclusion is reached by jointing together the MZR as given by stars (\cite[Kirby et al. 2013]{kir2013}) and PNe (\cite[Gon\c calves et al. 2014]{gon2014}) of the Local Group dwarf galaxies, as well as the H~{\sc ii} regions of the SDSS star-forming galaxies up to redshift 0.1 (\cite[Andrews \& Martini 2013]{anma2013}), as in Sect. 7. The highlights of the review are listed in a short, final section.

\section{PNe and their chemical abundances (O/H)}

Planetary nebulae are present in early- and late-type galaxies. Their emission lines are very bright and their chemistry provides information about the distant past of the galaxy (old- to intermediate-age stellar population). All these properties indicate that they are potentially good tracers of the chemical evolution in the LG. Otherwise similar to PNe, H~{\sc ii} regions are only present in late-type galaxies. H~{\sc ii} regions are usually brighter than PNe and are tracers of the present time ISM in a given galaxy. Together,  these two strong emission-line emitters give important constraints to the chemical evolution models, since in both the metallicity (gas-phase or nebular, O/H abundances) are derived in similar ways. For a review on the results for the chemistry of these two populations in nearby galaxies see, for instance, \cite[Magrini, Stanghellini \& Gon\c calves (2012)]{mag2012}.

In terms of chemical abundances, PNe can be divided in at least two categories, which in the Galaxy are defined mainly from their He/H abundances (\cite[Kingsburgh \& Barlow 1984]{kiba1984}). For instance, in the case of M33  (\cite[Magrini et al. 2009]{mag2009}), Type~I PNe are representing the recent past (they are \lq young', $<$~1~Gyr old progenitors, with masses  $>3$~M$_{\odot}$). Thus, Type-I PNe have  abundances more similar to those of the H~{\sc ii} regions.  Non-Type~I, on the other hand, probe the far past, with ages $\ge$~1~Gyr, and they are ideal to trace the M33's chemical evolution. The Type~I PNe definition depends on the metallicity, and at relatively low-Z (as in M33) it depends only on the N/O abundance ratio, regardless their He/H (log(N/O)$\ge$0.5; \cite[Dopita et al. 1991]{dop1991}). The relevant result to keep in mind here is: compared with the galaxy H~{\sc ii} regions, non-Type~I PNe are sufficiently old to trace the ISM chemical evolution.

When deriving the chemical abundances of PNe and H~{\sc ii} regions, the most relevant result are the oxygen abundances -- usually expressed as 12+log(O/H). This is so because the optical spectra of the photoionized nebulae contain most of the ions that contribute significantly to the total abundance of the oxygen (\cite[Stasinska 2004]{sta2004}; \cite[Delgado-Inglada 2017]{del2017}). Therefore we have oxygen as being the best indicator of the nebular abundances, either in H~{\sc ii} regions or PNe. If measured in the non-Type~I PN  population the O/H will  represent ISM more than 1~Gyr ago, at the time the PN progenitors were born. Of course this assumption will only be true if oxygen is only produced by stars that do not end-up as non-Type~I PNe. Otherwise, instead of representing the old ISM chemistry, the O/H would probe the nucleosynthesis of the progenitor stars. 

The analysis of the progenitor stars yields is done by understanding the third dredge-up effect (3du) on the stellar evolution, at different metallicities. We proceed to do this analysis in the next Section.

\section{Third dredge-up effects on the chemistry of PNe}

In PNe, the abundances of some elements are affected by the nucleosynthesis of the PN progenitors. These elements processed in the stellar inner zones can be dredged-up by convection to the envelope. In this way the  abundances  of  He,  C  and  N  are altered in surface layers during the giant branch and asymptotic giant branch (AGB) progenitor's evolution. As a matter of fact, a certain amount of oxygen and neon can  also be mixed in during the thermally pulsing phase of the AGB evolution (\cite[Kingsburgh \& Barlow 1984]{kiba1984}; \cite[P\'equignot et al. 2000]{peq2000}; \cite[Leisy \& Dennefeld 2006]{lede2000}). 

\cite[P\'equignot et al. (2000)]{peq2000} had demonstrated that the 3du only enrichs the O/H at low metallicities, and even in that case it is a rare process. \cite[Richer \& McCall (2007)]{rimc2007} confirmed this result in a number of nearby bright dIrr galaxies. However, a significant self-pollution of O in PN progenitors was found in some galaxies -- in Sex~A, \cite[Magrini et al. (2005)]{mag2005} and  \cite[Kniazev  et  al.  (2005)]{kni2005}; SMC, \cite[Leisy \& Dennefeld (2006)]{lede2000}; Fornax, \cite[Kniazev et al. (2007)]{kni2007}; Sagittarius, \cite[Kniazev et al. (2007)]{kni2007}. 

\begin{figure}
\begin{center}
 \includegraphics[width=4.8in]{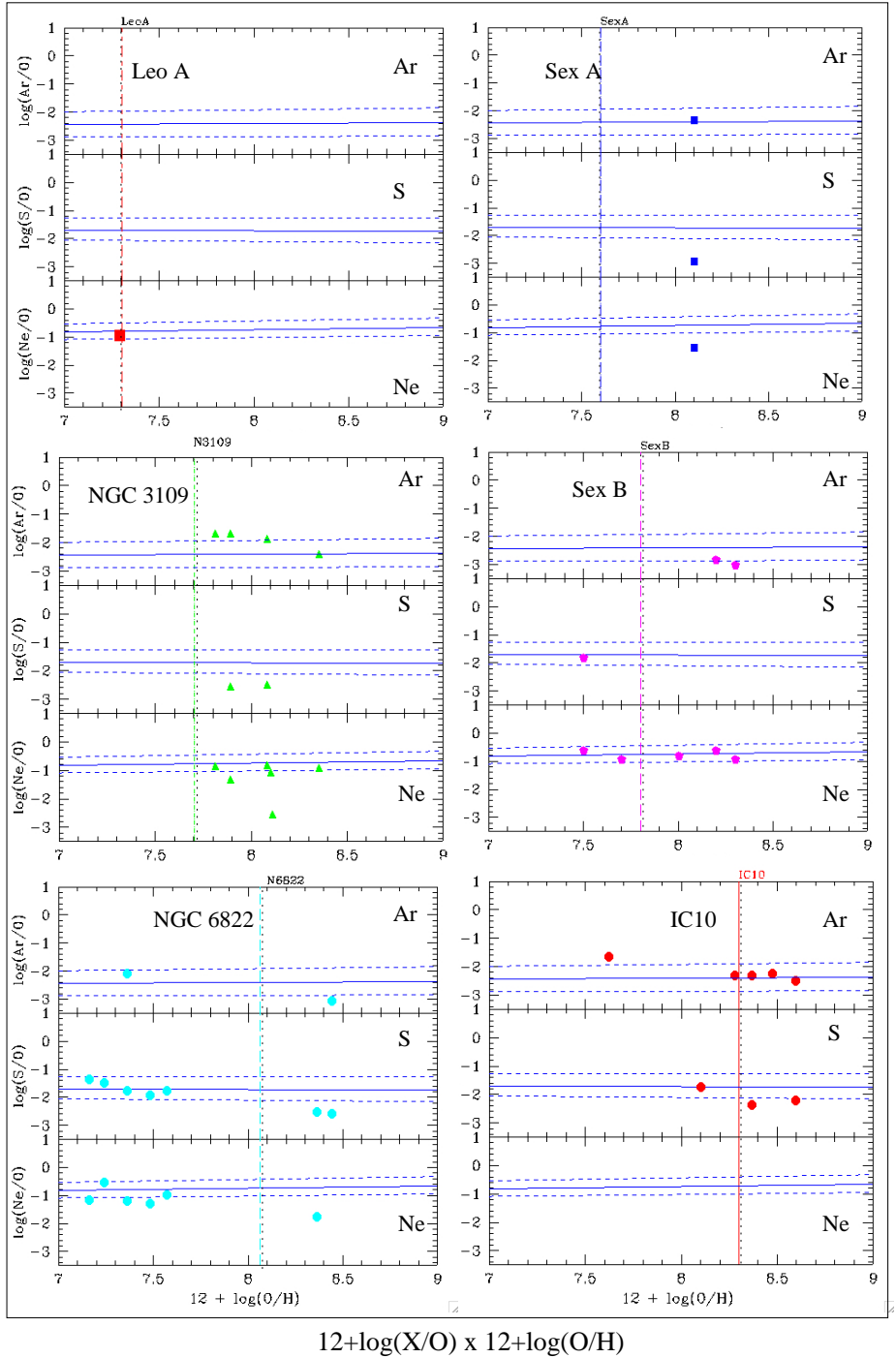} 
 \caption{The PNe abundance ratios log(X/O) vs. 12+log(O/H), with X given in each panel, along with the identification of the dIrr galaxies. The sources of data are as follow. 
In the left, top to bottom panels we show: Leo~A (van Zee et al. 2006); NGC~3109 (Pe\~na et al. 2007) and NGC~6822 (Hern\'andez-Mart\'inez et al. 2009). The right, top to bottom panels show: Sex~A (Magrini et al. 2005);  Sex~B (Kniazev et al. 2005; Magrini et al. 2005)  and IC~10 (Magrini \& Gon\c calves 2009). The quasi-horizontal lines show the abundance line ratios given by H~{\sc ii} regions in a sample of BCGs, as compiled by Izotov et al. (2006). The vertical lines represent the 12+log(O/H) of the  H~{\sc ii} regions of each dIrr galaxy.  
}
   \label{f3du}
\end{center}
\end{figure}

The possible presence of the 3du altering the O/H in the PNe of a number of nearby dIrr was studied by \cite[Magrini \& Gon\c calves (2009)]{mago2009}, and is shown in Figure~\ref{f3du}. dSph are not included in the plot, since
they do not have H~{\sc ii} regions and the comparison with the present-time ISM is not possible. The figure plots the PNe abundance ratios as a function of the O/H from H~{\sc ii} regions (vertical lines). The same PNe ratios are also contrasted with a sample of BCGs (blue compact galaxies, the \lq horizontal' lines). This is a relevant comparison because in H~{\sc ii} regions the oxygen observed has been produced by the same massive 
stars that produced the $\alpha$-process elements neon, sulphur and argon. Therefore,  log(Ne/O), log(S/O) and log(Ar/O) should be constant and show no dependence on the oxygen abundance. In fact, in 
the three diagrams of Figure~\ref{f3du}, most of the PNe are distributed along the line that defines the BCGs abundance ratios of H~{\sc ii} regions, therefore showing no trends with respect to the oxygen abundance. The oxygen production in these diagrams should have two
effects: (i) to place the PNe points well ahead of the mean abundance
in the ISM (so well ahead of the vertical lines), and (ii) to depress the Ne/O, S/O and Ar/O ratios, unless
the production of Ne, S and Ar also occurs. In a conservative view,
we expect that the presence of both effects would probe the
3du occurrence. Only a few PNe in Figure~\ref{f3du}, those of Sex~A and NGC~3109,   full-fill both requests. Also note from these plots that the 3du contribution is important only for galaxies with the lowest metallicities. 

As a matter of fact, the plots in Figure~\ref{f3du} are also giving us a lower limit above of which no third dredge-up effect is seen in the nearby dwarf irregulars. This limit is 12+log(O/H)$\leq$7.7 (\cite[Magrini \& Gon\c calves 2009]{mago2009}). 

\cite[Flores-Dur\'an et al. (2017)]{fldu2017} have recently shown that the PNe of NGC~3109 are not only enriched in O  
(\cite[Pe\~na et al. 2007]{pena2007}), but they are also enriched in Ne. In this galaxy the average O abundance of the H~{\sc ii} regions is 12+log(O/H)=7.74, and significantly higher (by 0.43dex) in PNe. Ne/H are ~3$\times$ higher in the PNe than in the H~{\sc ii} regions of NGC~3109. Is is not a coincidence that the ISM abundance of this dIrr is the same as the lower limit for the oxygen 3du determined years before by \cite[Magrini \& Gon\c calves (2009)]{mago2009}. NGC~3109 is one of the two lowest Z galaxies included in Magrini \& Gon\c calves (2009) and in Figure~\ref{f3du}.

\section{PNe vs. H~{\sc ii} regions abundance gradients in nearby galaxies}

\begin{table}
  \begin{center}
  \caption{Some global results of chemical abundance distributions given by PNe for S and dSph and also by H~{\sc ii} regions for the dIrr galaxies.}
  \label{tab1}
 {\scriptsize
  \begin{tabular}{|l|c|c|c|c|}\hline 
{\bf Spirals} & {\bf Type} & {\bf 12+log(O/H)}$_{PNe}$ & {\bf Gradient/Comments} & {\bf Reference} \\ \hline
MW		& S	& 8.70	&Flat -- up to ~28kpc				& Stanghellini et al. (2018)	\\ \\
M31		& S	& 8.62	&Flat -- beyond ~20kpc				& Balick et al. (2013)	\\ \\
NGC~300	& S	& 8.57	&Flat -- up to ~5kpc					& Stasinska et al. (2013)	\\ \\
M81		& S	& Average 8.54	& Negative					& Stanghellini et al. (2010)	\\ \\	
M33		& S	& Average 8.29	& Negative -- up to ~8kpc	& Magrini et al. (2009)	\\ \\
LMC		&Irr& 8.27	& Non-homogeneous                    & Leisy \& Dennefeld (2006);	\\
   		&   &    	& abund. distribution                &  Kwitter \& Henry (2012)	\\ \\
SMC 		&Irr& 8.04	& Non-homogeneous                    & Leisy \& Dennefeld (2006); 	\\
    		&   &    	& abund. distribution                &  Kwitter \& Henry (2012)	\\
\hline
{\bf Irregulars} & {\bf Type} & {\bf 12+log(O/H)$_{PNe/HIIr}$} & {\bf Gradient/Comments} & {\bf Reference} \\ \hline
IC~10		& dIrr	& 8.30/8.30	& Non-homogeneous				& Magrini \& Gon\c calves (2009)	\\
    			&       &    	    & abund. distribution            &  	\\ \\
NGC~6822		& dIrr	& 8.06/8.06	& Flat -- up to ~2kpc				& Hern\'andez-Mart\'\i nez et al. (2009)	\\ \\
NGC~3109		& dIrr	& 8.17/7.74	& Flat -- up to ~2.5kpc			& Flores-Dur\'an et al. (2017)	\\ \\
Sextans~A	& dIrr	& 8.00/7.60	& --    							& Magrini et al. (2005)	\\ \\	
Sextans~B	& dIrr	& 8.00/7.80	& --    							& Magrini et al. (2005)	\\ \\
\hline
{\bf Spheroidals} & {\bf Type} & {\bf 12+log(O/H)$_{PNe}$} & {\bf Gradient/Comments} & {\bf Reference} \\ \hline
Sagittarius		& dSph	& 8.40+-015	& Steep Z-age 				& Zijlstra et al. (2006)	\\
    			&       &    	    & gradient            &  	\\ \\
Fornax	& dSph	& 8.28$\pm$0.20	&  --			& Kniazev et al. (2007)	\\ \\
NGC~185	& dSph	& 8.20$\pm$0.30	&  -- 		& Gon\c calves et al. (2012)	\\ \\
NGC~205 & dSph	& 8.08$\pm$0.28	& --    							& Gon\c calves et al. (2014)	\\ \\	
NGC~147	& dSph	& 8.06$\pm$0.10	& Homogeneous     							& Gon\c calves et al. (2007)	\\ 
    			&       &    	    & abund. distribution            &  	\\ 
\hline
  \end{tabular}
  }
 \end{center}
 \end{table}

\begin{figure}
\begin{center}
 \includegraphics[width=4.25in]{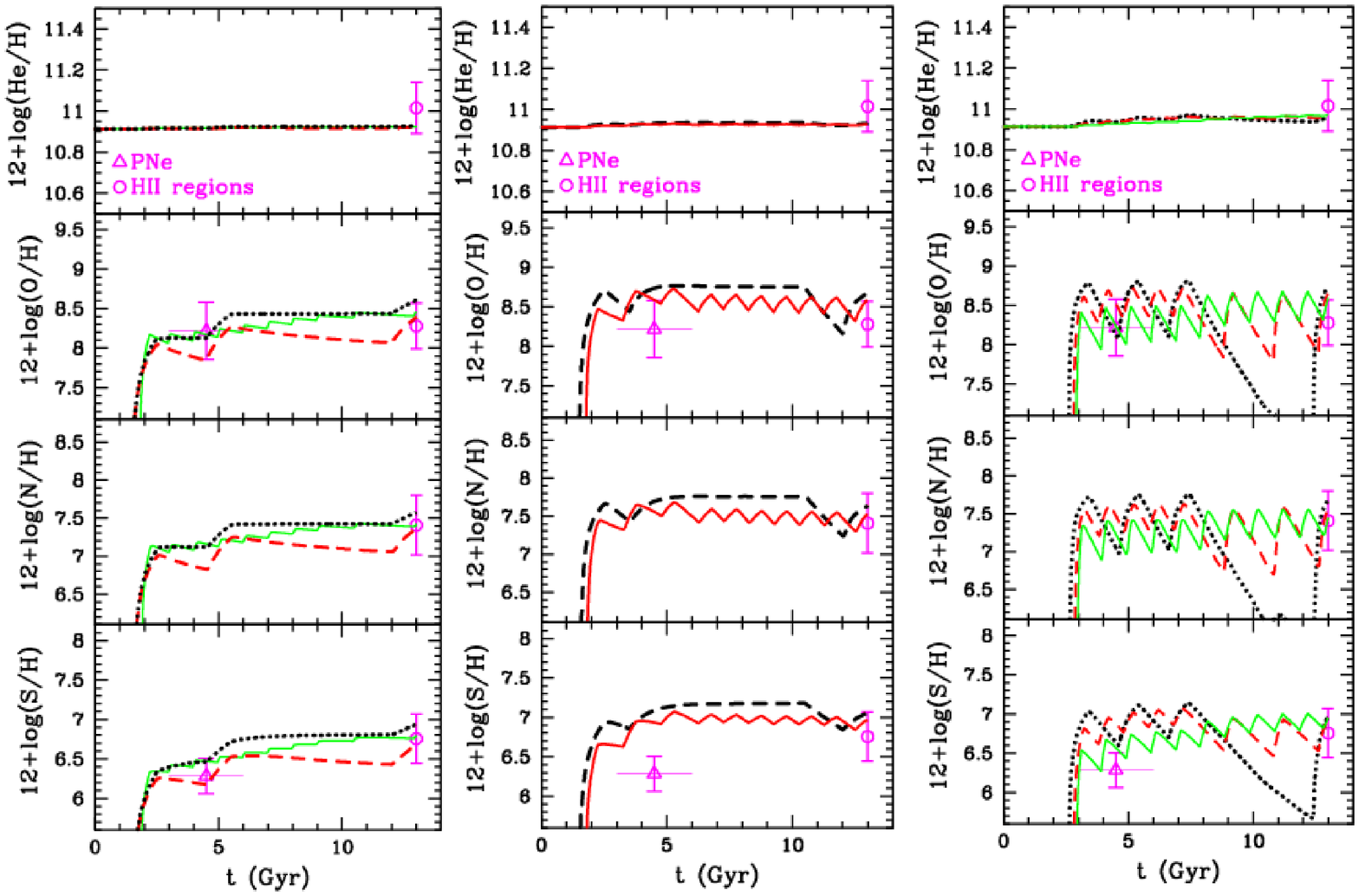} 
 \includegraphics[width=4.25in]{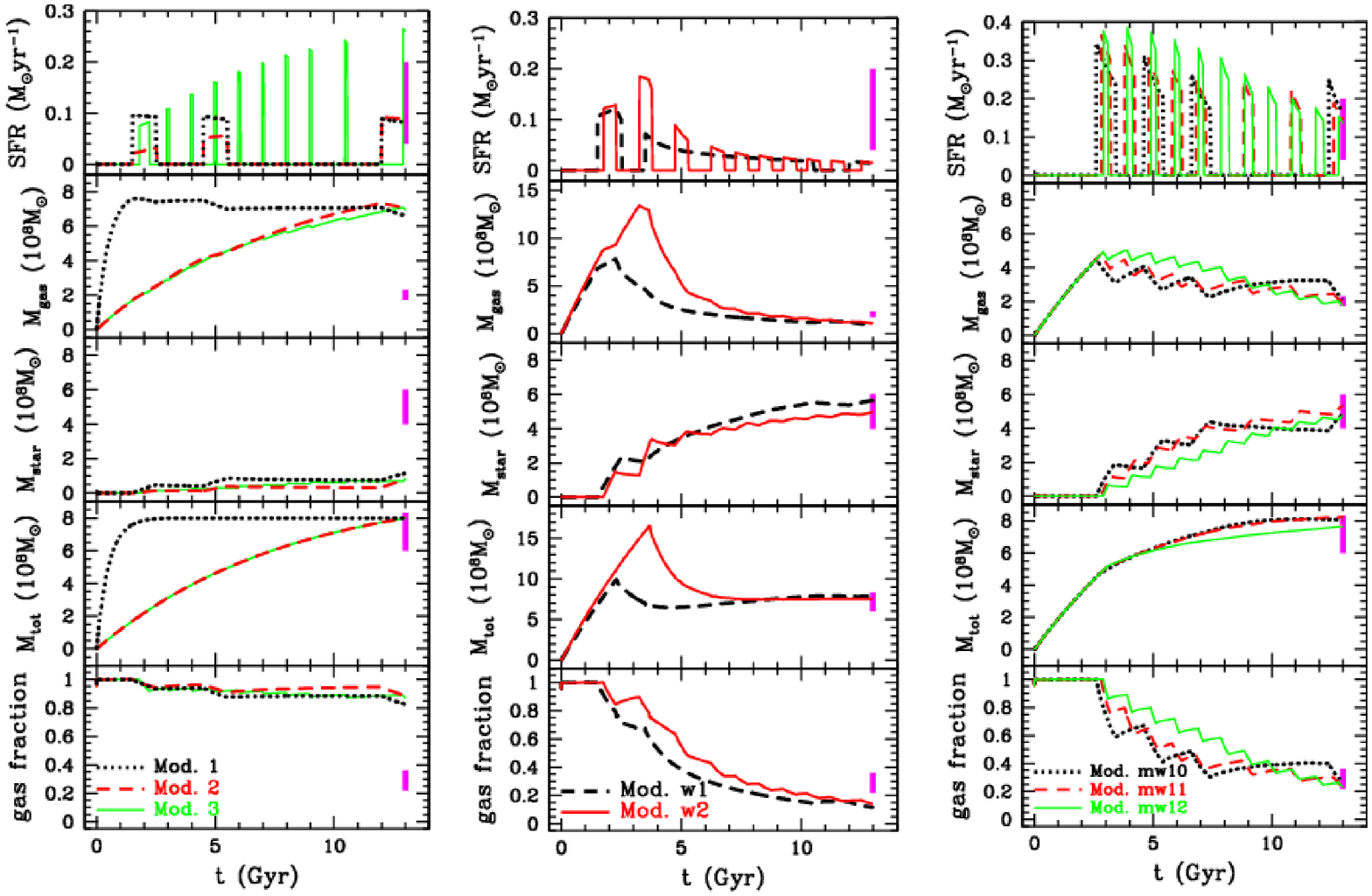} 
 \caption{Time evolutions predicted by the models for IC~10 in Yin et al. (2011). 
The magenta symbols stand for the PNe (triangles) and H~{\sc ii} (circles) chemical abundance ratios. The magenta bars at 13~Gyr represent the range of the observational data (see Table~1 of Yin et al. 2010). The dotted, dashed and continuous lines represent models with no winds (left panels), with normal winds (central panels) and with metal-enhanced winds (right panels), with different assumptions concerning the number of star-formation bursts and infall timescales. The evolution of He, O, N and S abundance ratios, as well as that of the total SFR, total gas mass, total 
 stellar mass, total baryonic mass and gas fraction are shown on the 9 panels. 
 Only the models in the right panels, models with metal-enhanced winds, can account for all the observational constraints (Yin et al. 2011). 
 }
   \label{fCEM1}
\end{center}
\end{figure}

\begin{figure}
\begin{center}
 \includegraphics[width=4.25in]{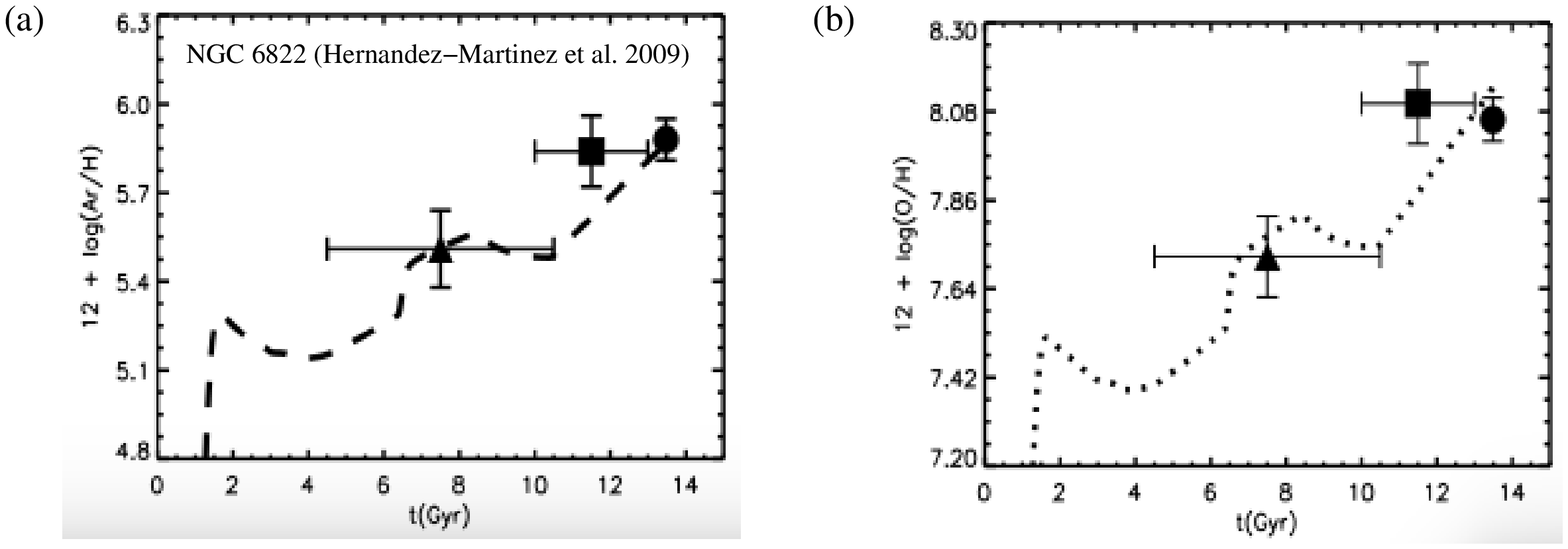} 
 \caption{Argon and oxygen time evolutions predicted by Hern\'andez-Mart\'inez et al. (2009) chemical evolution models for NGC~6822. The 12+log(Ar/H) (a) and 12+log(O/H) (b) evolution, as constrained by H~{\sc ii} regions (circles), young PN (boxes) and old PN (triangles) populations.}
   \label{fCEM2}
\end{center}
\end{figure}

The general discussion of metallicity gradients in irregular galaxies seems to indicate that most galaxies have  spatially homogeneous composition (e.g., Kobulnicky \& Skillman 1997; Croxall et al. 2009; Haurberg et al. 2013; Lagos \& Papaderos 2013; Hosek et al. 2014; Patrick et al. 2015). Also see Venn et al. (2004), Lee et al. (2006) and Annibali et al. (2015), for exceptions to this trend. Following Pilyugin et al. (2014, 2015) galaxies with steep (flat) inner surface brightness profile usually present noticeable (shallower or null) radial abundance gradients. This behaviour might be related to the presence of radial mixing of gas, a process that takes place more evidently in galaxies with flat surface density inner profiles. 
In Table~1 we compile the abundance distributions for a few spiral, dwarf irregular and dwarf spheroidal galaxies, as given by PNe (S and dSph) and also by H~{\sc ii} regions (dIrr). Different behaviours -- flat and negative gradients, as well as homogeneous and non-homogeneous abundances distributions -- can be seen.

In Magrini et al. (2016), late- (M33 and NGC~300) and the early-type (M31 and M81) spirals are compared. Chemical abundance gradients as given by non-Type~I PNe and H~{\sc ii} regions are considered. It is found that, at a given galactocentric distance, the H~{\sc ii} regions are more metal rich than the PNe. This is equivalent to say that all these galaxies are still being enriched in metals. Moreover, what these data show is that the gradients do not allow us to distinguish if they represent slight flattening or slight steepening with time. However, stronger gradients cannot be excluded from the birth of the non-Type~I PNe (~5 Gyr ago) to the present (see Magrini et al. 2016, Figures 6 to 9). 

Now turning our attention to the dIrr galaxies in Table~1, we note that not only the PNe average abundances but also the H~{\sc ii} regions ones are given. With this we want to highlight that  usually the chemical evolution model constraints are the O/H from H~{\sc ii} regions, (O/H)$_{HIIr}$, and the [Fe/H] from old stars. Recently, however, the PNe O/H, (O/H)$_{PNe}$, is also being used. And, even more recently, PNe are divided into the young (Type~I PNe) and the old ones (non-Type~I PNe). The abundances of the former are similar to the (O/H)$_{HIIr}$ abundances, while those of the latter roughly corresponds to the [Fe/H] abundances of old stars. Therefore chemical evolution models for dIrr galaxies can be performed without any O/Fe conversion!

Indeed, the chemical evolution model of the dIrr galaxy IC~10 was performed 
in such a way by \cite[Yin et al. (2010)]{yin2010}. This model takes advantage of the detailed abundances made available by \cite[Magrini \& Gon\c calves (2009)]{mago2009} and follows the recipes of the chemical evolution models for dwarf irregulars (\cite[Yin et al. 2011]{yin2011}). The main ingredients of the model are as follows. i) Detailed  metallicity-dependent  stellar  yields  for  both massive and low-to-intermediate mass stars. ii) Feedback from SNe. iii) Stellar winds that follow the development of galactic winds. And iv) several regimes of star formation -- bursting mode with short and long bursts, and a continuous but low star-formation rate (SFR). To reach the best-fitting model, three cases were explored. Models with no winds (no feedback), as in   Figure~\ref{fCEM1} left panels. Models with normal winds, namely all the gas is lost at the same rate  (middle panels of Figure~\ref{fCEM1}). And finally models with metal-enhanced winds, where 
metals are lost preferentially relative to H and He (right panels of Figure~\ref{fCEM1}). This figure clearly shows that galactic evolution models without winds predict too high gas fraction (left panels). If, on the other hand, normal wind are considered, the result is a very low SFR at the present (middle panels). The models that work are those with metal enhanced wind, on which a continuous wind follows the first SF burst, and carries away mostly the metals (right panels; \cite[Yin et al. 2010]{yin2010}). 

Chemical evolution models tailored to understand the ISM of dIrr galaxies were also published by \cite[Hern\'andez-Mart\'\i nez et al. (2009, 2011)]{}. In Figure~\ref{fCEM2}, abundances from H~{\sc ii} regions (circles, at about 13~Gyr) and those from the PN populations of two ages (young PNe, boxes; and old PNe, triangles) were used to constrain the evolution model of NGC~6822. Though a definitive answer is not reached in the paper, \cite[Hern\'andez-Mart\'\i nez et al.  (2011)]{hen2011} also addressed the most relevant problem of the nebular abundances discrepancy.  The problem is due to the use of the collisionally-excited lines (CELs) or the optical recombination lines (ORL) for the abundance determination (\cite[Wesson et al. 2018]{wes18}). This is a promising way of shedding light to this problem, since a number of other than ISM chemical evolution constraints are present in the galactic evolution models. Therefore, if we can differentiate between the set of abundances (CELS or ORLs) that account for the total mass, the gas and stars mass fractions and the SFRs of the evolution models, we can also solve this long standing problem. 

Perhaps somewhat surprisingly, Flores-Dur\'an et al. (2017) detailed model of another dIrr galaxy, NGC~3109, have made very clear that oxygen and neon are  enriched in its PN population, as compared to in the H~{\sc ii} regions. So in this galaxy the 3du is stronger for Ne than for O. This behaviour is hard to explain by classical stellar evolution. The kinematics of the PNe, the blue supergiant stars (BSG), the H~{\sc ii} regions and that of the H~{\sc i} gas in the galaxy are contrasted in this work. It is found that the PNe and BSG have similar kinematics distribution, whereas the H~{\sc ii} regions rotate in the same direction as the H~{\sc i} disk (Flores-Dur\'an et al. 2017). 
 This behaviour is interpreted as an indication that younger and intermediate-age objects of NGC~3109 belong to different populations, probably located in the galaxy central stellar bar.

With the focus on other aspects of the dwarf spheroidals evolution, \cite[Otsuka et al. (2011]{ots2011}; Sagittarius) and \cite[Martins et al. (2012]{mar2012}; NGC~185) actually take advantage of the PN abundances in their studies. The case of the latter is indeed curious, since to NGC~185 it was not only attributed the classification as a dSph, but also as a Seytfert galaxy.  Although its emission lines place it in the category of Seyferts, the galaxy was not detected either in radio surveys or in 6 and 21~cm, or X-ray observations, which probably indicates that the Seyfert-like line ratios may be produced by stellar processes. The strong emission-lines from the post-AGB population could perhaps fake a Seyfert galaxy. \cite[Martins et al. (2012)]{mar2012} had, instead, shown that a combination of PNe plus supernova remnants is the responsible for the mimic. (O/H) from PNe are used in the chemical evolution models of this work to predict the supernova rates and to support the idea that supernova remnants should then be present in NGC~185, as in fact observed (\cite[Gon\c calves et al. 2012]{gon2012}).

\section{Mass-metallicity relation for star forming galaxies}

\begin{figure}
\begin{center}
 \includegraphics[width=5.25in]{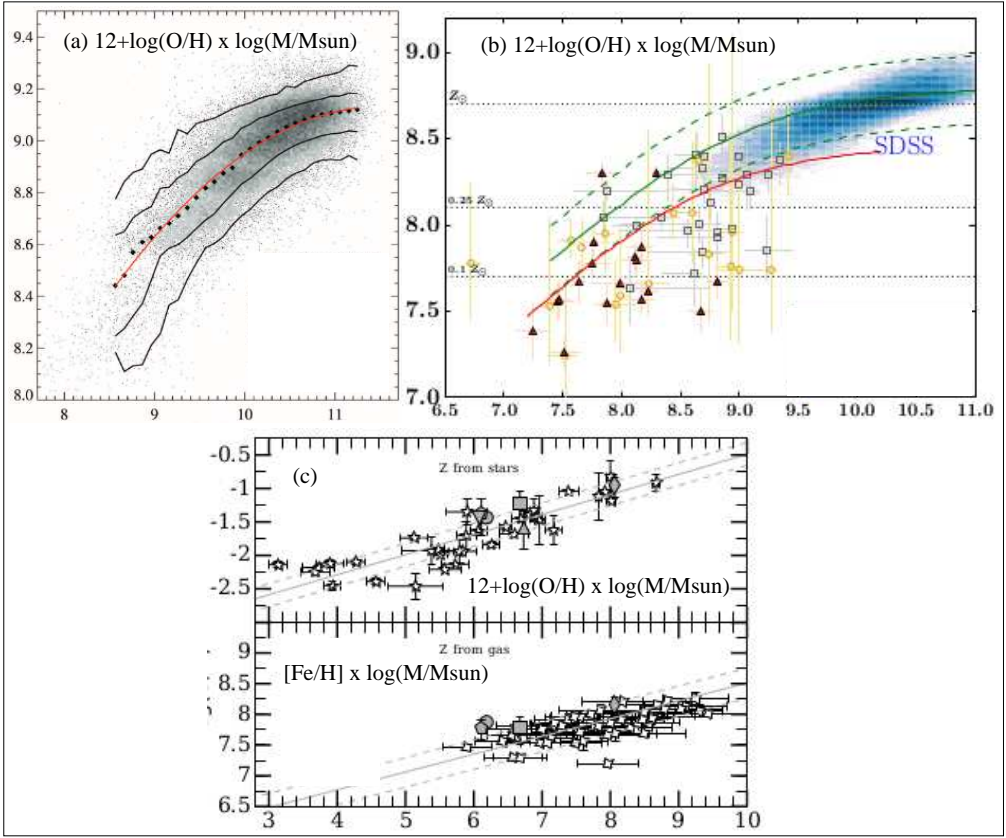} 
 \caption{Tree examples of MZR, for normal and dwarf galaxies. {\it (a)} Gas-phase oxygen abundance for $\sim$53,400 SF galaxies in the SDSS vs. stellar masses. The black dots are the median in bins of 0.1 dex in mass, while the red line shows a polynomial fit to the data (Tremonti et al. 2004). {\it (b)} MZR based on the 0.1 $\leq$ z $\leq$ 0.9 VIMOS Ultra Deep Survey SF galaxies (filled red triangles) to be compared with the O/H (empty circles and squares) compiled by Ly et al. (2014, 2015, respectively). \cite[Andrews \& Martini (2013)]{anma2013} 
MZR and its $\pm$1$\sigma$ uncertainty are represented by the green continuous and dashed lines, while the lower red continuous line indicates the MZR from Ly et al. (2016), from Fig.~19 in  
Calabr\`o et al. (2017). {\it (c)} The MZR from the spectroscopy of individual stars
(upper panel; fit by Kirby et al. 2013) and from gas-phase metallicities (lower panel; fit by Berg et al. 2012), in comparison with the isolated Local Group dwarfs in (Hidalgo 2017).
}
   \label{fMZR}
\end{center}
\end{figure}

The mass-metallicity relation (MZR) as described by \cite[Tremonti et al. (2004)]{tre2004} is well known as one of the fundamental relations of galaxies. It was built with the metallicity 12+log(O/H) from 53,400 SDSS SF galaxies with redshift (z) up to $\sim$~0.1, as a function of the stellar mass. In this relation the stellar mass (M$_*$) reflects the amount of gas locked up into stars, while the metallicity reflects the 
gas reprocessed by stars and any gas exchange between the galaxy and its environment. It is then impressive that Tremonti et al. (2004) found a tight  correlation between M$_*$ and Z.  Moreover, it spans over 3 orders of magnitude in  M$_*$
 (8.5 $\leq$~log(M$_*$/M$_{\odot}$)~$\leq$ 11.5) and a factor of 10 in Z (8.2$\leq$12+log(O/H)$\leq$ 9.2), as can be seen in Figure~\ref{fMZR}, panel (a). This relation in fact represents the interplay of different effects, as, for instance, star formation efficiency, stellar winds, in- and out-flows, and of course the depth of the potential wells. Keeping in mind all these phenomena -- each of them particularly relevant for dwarf galaxies -- it is then mandatory to explore to which extend the relation holds for nearby dwarf galaxies. 

The subject of dwarf galaxy's MZR has been actively investigated lately, and a few conclusions in these works are mentioned here (Fig.~\ref{fMZR}-\ref{f7ul}). Two recent examples can be seen by inspecting Figure~\ref{fMZR}, which shows together the MZR as given by Tremonti et al. (2004), 
\cite[Calabr\`o et al. (2017)]{cal2017} and \cite[Hidalgo (2017)]{hid2017}. The MZR in panel (b) was derived to prove its low-mass end in the 0.1 $\leq$ z $\leq$ 0.9 bright star-forming galaxies of the VIMOS Ultra Deep Survey (Calabr\`o et al. 2017). The latter authors emphasize that the scatter 
for the lower masses can be explained by variations in the specific SFRs and the gas fractions of the galaxies. 
In panel (c) of Figure~\ref{fMZR}, on the other hand, Hidalgo (2017) built the dwarf galaxies MZR to investigate the time dependence of the relation. 
The result is that, unless the SFR is taken as an additional parameter in the MZR, no time evolution is found. 
The main Hidalgo's conclusion is that, for a given M$_{*}$, Z decreases with the SFR, also clearly emphasizing the role of the SFR on the mass-metallicity relation.

Either for dwarf or higher mass galaxies, the MZR is usually defined for the SF galaxies, since they have strong emission-lines coming from the H~{\sc ii} regions. Note, however, that in  Figure~\ref{fMZR}, panel (c), the relation is also presented for the stellar ([Fe/H]) metallicity, as opposite to the 12+log(O/H) gas-phase one. Following Mateo~(1998) and earlier discussions in this review, the oxygen abundances can be converted into iron metallicity. But the assumptions about the [O/Fe] are very uncertain, since this  ratio depends on the SFH of the galaxies (Gilmore \& Wyse 1991). As so, the conversion between the these two Z indicators  should preferentially be avoided. Therefore, if in addition to study the MZR of the SF galaxies we also want to understand the relation in the quiescent ones, we need to adopt a metallicity indicator present in SF as well as quiescent galaxies. PNe can provide such a safe Z indicator in both types of galaxies. Thus, PNe allow us to add to the SF galaxies mass-metallicity relation the behaviour of the non-star-forming ones, without the need of the O to Fe conversion, as we show in the next section.

\section{The stellar and PNe luminosity-metallicity relation}

\begin{figure}
\begin{center}
 \includegraphics[width=5.25in]{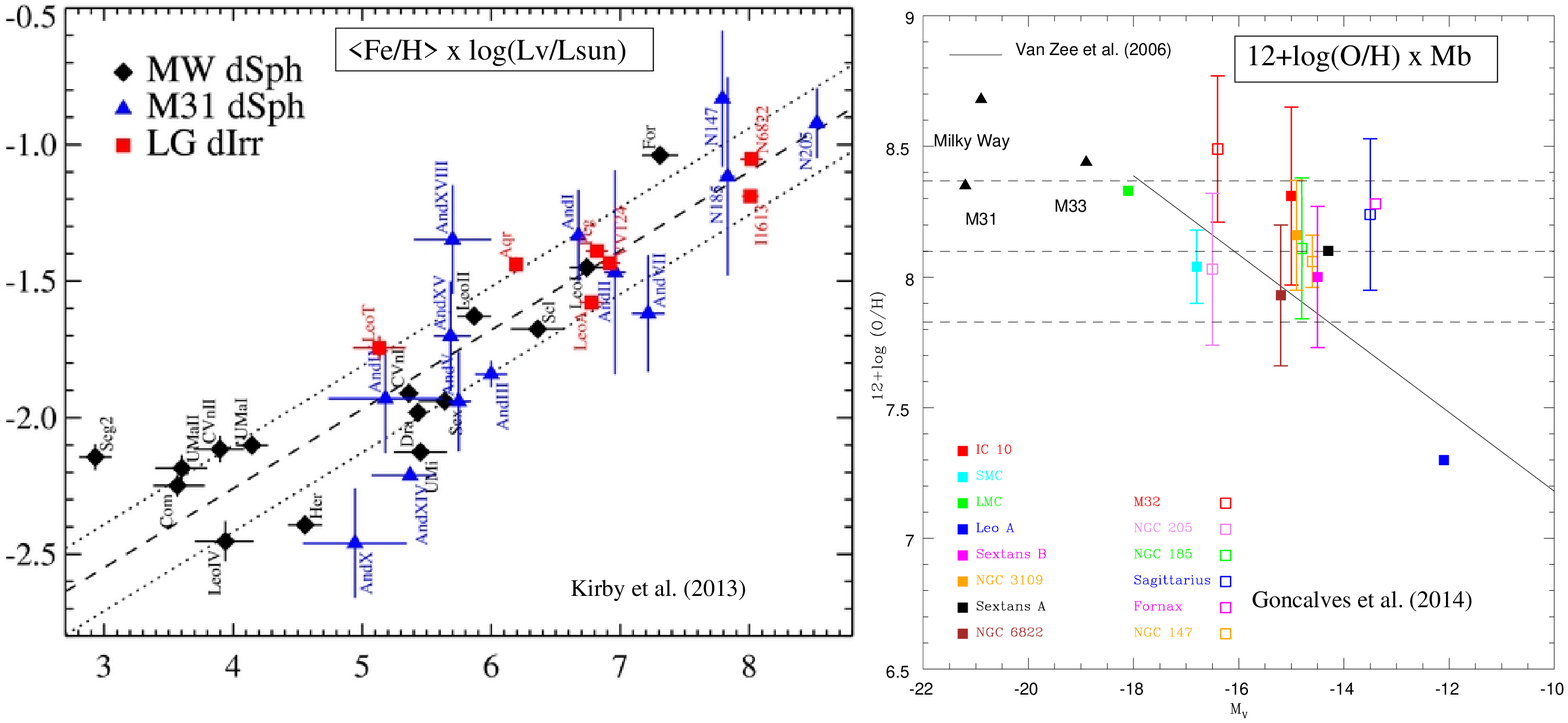} 
 \caption{The stellar (Kirby et al. 2013) and planetary nebulae (Gon\c calves et al. 2014)
 luminosity-metallicity relations for the Local Group dwarf galaxies. {\it Left}: Diamonds, squares and triangles represent the average [Fe/H] from the spectroscopy of individual stars in the MW dSph satellites, dIrr galaxies and M31 dSph satellites, respectively. The dashed line shows the least-squares, where the intercept is calculated at 10$^6$ L$_{\odot}$, while the dotted line shows the rms about the best fit (see Eq.~3 of Kirby et al. 2013).  {\it Right}: The PNe LZR is given as 12+log(O/H) vs. magnitude (M$_B$). Filled symbols represent the dIrr, while the dSph galaxies are shown with empty symbols. The standard deviation of the abundances in a given galaxy is also plotted, with the exception of the cases on which only one PN has its oxygen abundance determined. The continuous line represents the LZR from H~{\sc ii} regions of nearby (D $\le$ 5~Mpc) galaxies with magnitudes fainter than -18 (van Zee et al., 2006).  The dashed lines represent the mean value and standard deviation (8.098$\pm$0.27) for the PNe of all the dwarf galaxies in the plot. 
The references for 12+log(O/H) are giving in Gon\c calves et al. (2014), while the L (M$_B$) of the dwarf galaxies are from the Mateo~(1998), with the exception of those for the LMC and SMC that are from Lee et al.~(2003). M$_B$ of the Galaxy, M31 and M33 are from van den Bergh~(2007).
}
   \label{fLZR}
\end{center}
\end{figure}

Before to study the MZR of the LG dwarf galaxies, we also want to establish the luminosity-metallicity relation (LZR) for the dSph and dIrr Local Group galaxies, in a fully homogeneous way. We strongly believe that this kind of study is the key to disentangle possible differences between star-forming and non-star-forming dwarf galaxies, as claimed long ago by Skillman et al. (1989) and Kormendy \& Djorgovski (1989). Is there an evolutionary correlation among these two types of dwarf galaxies? Could the dSph galaxies be formed through the removal of the gas in dwarf irregulars, either through ram pressure stripping, supernova driven winds or star formation fading? Would we see such an evolution in the luminosity- or mass-metallicity relations?

Already in Richer \& McCall~(1995) PNe were proposed as an ideal population to measure the 
metallicity of the dwarf spheroidals and irregulars. These authors, and later also Gon\c calves et al. (2007), actually successfully performed such an exercise. Both groups found that the dSph's LZR presented a significant offset from the dIrr relation, with PNe in dSph galaxies having higher O/H than those of the dIrr ones. Gon\c calves et al. (2007) pointed out that  the LZR of dSph galaxies did not exclude their formation from old dwarf irregulars, but it did exclude their formation from the present time dIrr galaxies, since the differences between their metallicities were already present in the old PN populations of both types. The LZR offset, then, would indicate a faster enrichment of dSph galaxies. The different SFHs for these two types of galaxies were also discussed in Grebel (2005).   
 
Grebel et al. (2003) combined photometric and spectroscopic red giant branch (RGB) star abundances, and built a stellar LZR. The above mentioned offset also was found, and it was shown that it persisted when the comparison was restricted to the galaxies' old stellar populations. Among these galaxies Grebel et al. (2003; also see Koleva et al. 2013) identified ‘‘transition-type dwarfs’’. These transition objects have   mixed dIrr/dSph morphologies, low stellar masses, low angular momentum, and dIrr contents of at most a few 10$^6$~M$_{\odot}$, which then closely resemble dSph galaxies,  if their gas were removed. The obvious conclusion was that these transition objects were likely the dSph progenitors. 

Surprisingly, in recent works based on stellar metallicities on which, more consistently, only spectroscopic stellar metallicities are used (Lee et al. 2008; Kirby et al. 2013), the above offset was seriously questioned. The caveat in Grebel et al. (2003) approach was the fact that the colors of red giants are subject to the age-metallicity degeneracy (Salaris \& Girardi 2005; Lianou et al. 2011). Lee et al. (2008) revised the RGB metallicities of the dIrr in Grebel et al. (2003) showing them to be 0.5 dex higher than previously published, with the consequence of vanishing completely the LZR offset! And, to definitely close this issue, Kirby et al. (2013) used spectroscopic Z of individual stars in seven gas-rich dIrr galaxies, and
found that dIrr obey the same LZR and MZR as the dSph galaxies. 

In Figure~\ref{fLZR} we compare Kirby et al. (2013) and Gon\c calves et al. (2014) LG dwarf galaxies LZRs, from purely spectroscopic stellar and PNe metallicities, respectively. Having no   photometric metallicities in this figure, the age-metallicity degeneracy of the stellar Z is not anymore an issue. It is evident from the Fig.~\ref{fLZR} that unlike the results just mentioned, no metallicity offset is found between the dIrr and dSph luminosity-metallicity relations! 

If in the case of the stellar relation the age-metallicity degeneracy of the photometrically obtained Z was the responsible for faking an offset, in the case of the PNe the poorly determined 12+log(O/H) can 
certainly be blamed. From 2007 to 2014, the O abundances of half of the galaxies in the plot were revisited, by re-observing the galaxies with 8-m class telescopes, as we detailed in Gon\c calves et al. 2014's paper.
Robust enough data were only made available more recently with spectrographs/telescopes able to provide deep optical spectra for the faint PN populations of the nearby dwarf galaxies. Also in the case of the PN populations LZR no offset is found anymore! Therefore, the important massage to learn here is that this kind of studies can only be carried out with high-quality data. Otherwise wrong conclusions will turn it even harder to understand the dwarf galaxies evolution. 

\section{PNe and the universality of the mass-metallicity relation}

The mass-metallicity relation given by the PN population in the LG dwarf galaxies is shown in Figure~\ref{fPNeMZR}. The first aspect to be noted from this plot is that, as in 
the case of the recent LZRs of Figure~\ref{fLZR}, dIrr and dSph galaxies are not segregated, and follow a common mass-metallicity relation. From the new MZR for the stellar content of the LG dwarf galaxies one also reaches the same conclusion (see, Figure~9 by Kirby et al. 2013). These authors, moreover, point out that their MZR is roughly continuous with the stellar mass-metallicity relation for galaxies as massive as M$\star$ = 10$^{12}$~M$_{\odot}$. 
 
Superposed to the planetary nebulae MZR, in the upper panel of Figure~\ref{fPNeMZR}, we also plot the least-squares fit obtained by Kirby et al. (2013) using the stellar metallicities. To this end, we added a constant to the Kirby's et al. (2013) interception of the Y-axis to match the nebular metallicity expressed in the PNe plot as 12+log(O/H):

\begin{figure}
\begin{center} 
 \includegraphics[width=4.05in]{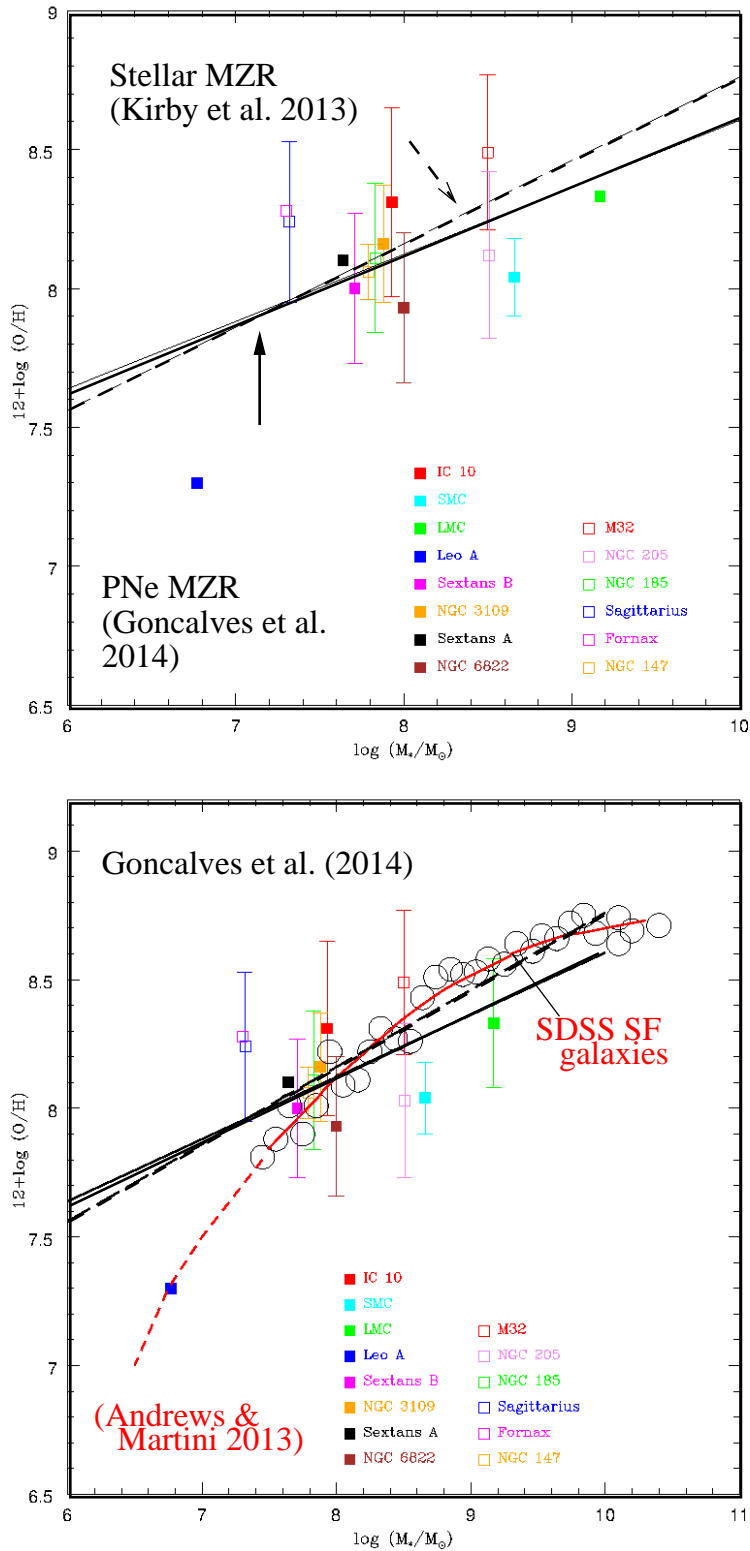} 
\caption{{\it Top:} The PNe mass-metallicity relation for the LG dwarf galaxies: showing 12+log(O/H) vs. 
   stellar mass. Filled symbols represent the dIrr while the dSph galaxies are represented by empty symbols. The standard 
   deviation of the abundances in a given galaxy is also plotted, except for the galaxies in which 
   only one PN has its oxygen abundance determined. References for the 12+log(O/H) can be found in  
   Gon\c calves et al. (2014). Data for stellar masses were taken from the compilation by McConnachie~(2012). 
   The dashed line represents the stellar MZR of Kirby et al. (2013) to which 
   a constant was added in order to match the oxygen metallicity, while the continuous line 
   is the mean least square fit of the PNe data. {\it Bottom}: The same as in the top panel, but also showing the mass-metallicity relation for M$_\star$ stacks in SDSS Data Release 7 (DR7; Abazajian et al. 2009) found by Andrews \& Martini (2013; empty circles). The metallicities in the SDSS relation were obtained via the direct-method. 
   The red 
   solid curve shows the asymptotic logarithmic fit to the SDSS direct method measurements. The dashed red curve is an extrapolation of the fit, up to the last point given by the lower luminosity dwarf galaxy of the sample. 
}
   \label{fPNeMZR}
\end{center}
\end{figure}

\begin{equation}
12 + log~({O \over H})=(-1.69-1.8+A)+0.3\times log~({M\star \over M_{\odot}})
\end{equation}
where -1.69 is the original intercept from the fit of Kirby et al. (2013), obtained using log~(M$\star\times10^6$M$_{\odot}$),  
$-$1.8 is the constant that takes into account that our abscissa is expressed in log(M$\star$/M$_{\odot}$)
and A=9.25 is the constant needed to match the data.  
The choice of A=9.25 dex translates the Kirby et al. (2013) relation to the
region of Figure~\ref{fPNeMZR} occupied by the PNe data. Note, however, that while the slope is relevant for the 
comparison of the PNe data and the other relations in Figure~\ref{fPNeMZR} (upper and bottom panels), the zero point is of no importance. 
The least mean square fit to the PNe data gives the following relation: 

\begin{equation}
12 + log~({O \over H})=6.19 + 0.24\times log~({M\star \over M_{\odot}}). 
\end{equation}

The MZR with oxygen PN abundances seems to deviate from the stellar one only for masses as smaller as that 
of Leo~A ($\sim$10$^7$~M$_{\odot}$). However, if the comparison with the mass-metallicity relation for the SF galaxies in the SDSS, and its extrapolation, are considered, Leo~A is not anymore an outlier of the MZRs. Andrews \& Martini (2013) adopted the direct method oxygen abundances from the stacked spectra of $\sim$200,000~SDSS\footnote{SDSS Data Release 7 by Abazajian et al. (2009).} star-forming galaxies, in bins of  0.1 dex in stellar mass, to build the relation we present in the bottom panel of Figure~\ref{fPNeMZR}. The stacked data (open circles), the continuous line fitting these data and the dashed-line extrapolation of this fit are evident in the figure. The quality of Andrews \& Martini (2013) relation is based on the fact thet they measured the [O~{\sc iii}], [O~{\sc ii}], [N~{\sc ii}] and [S~{\sc ii}] electron temperatures of  stacked spectra, and were able to derive the best possible O/H from the SDSS galaxies. 

Even though Gon\c calves et al. (2014) discuss arguments questioning the positions of Fornax and Sgr dward galaxies in the MZR, one can easily note that the oxygen abundances of  PNe trace the same metallicity as the H~{\sc ii} regions in the SDSS. Particularly, no offset from the average 
metallicity in each bin of stellar mass can be spotted. Moreover, it is  worth noticing that the dIrr  galaxy Leo~A is perfectly consistent with the extrapolation of the asymptotic 
logarithmic fit of the SDSS mass-metallicity relation (the dashed line that reaches stellar masses as smaller as log(M$\star$/M$_{\odot})\sim$6.6) shown in Figure~\ref{fPNeMZR} bottom panel. 

We recognize that there is a non-negligible scatter among the data points based upon the chemical composition from PNe. Even though, the present results (Gon\c calves et al.~2014) are compatible with a unique mass-metallicity relation for the dSph and dIrr LG galaxies. Also, the bottom panel of Figure~\ref{fPNeMZR} strongly suggests the universality of the MZR for both dIrr and dShp galaxies. 

Having learned with the previous experience, we are very conservative in summarizing these results as follows. {\em Presently neither the data nor the MZR are known well enough to categorically affirm that dwarf irregulars and dwarf spheroidals follow the same mass-metallicity relation, though they are now fully compatible with this possibility. This was not the case previously!}

\subsection{Further results on the MZR for dwarf galaxies}
 
\begin{figure}
\begin{center}
 \includegraphics[width=5.0in]{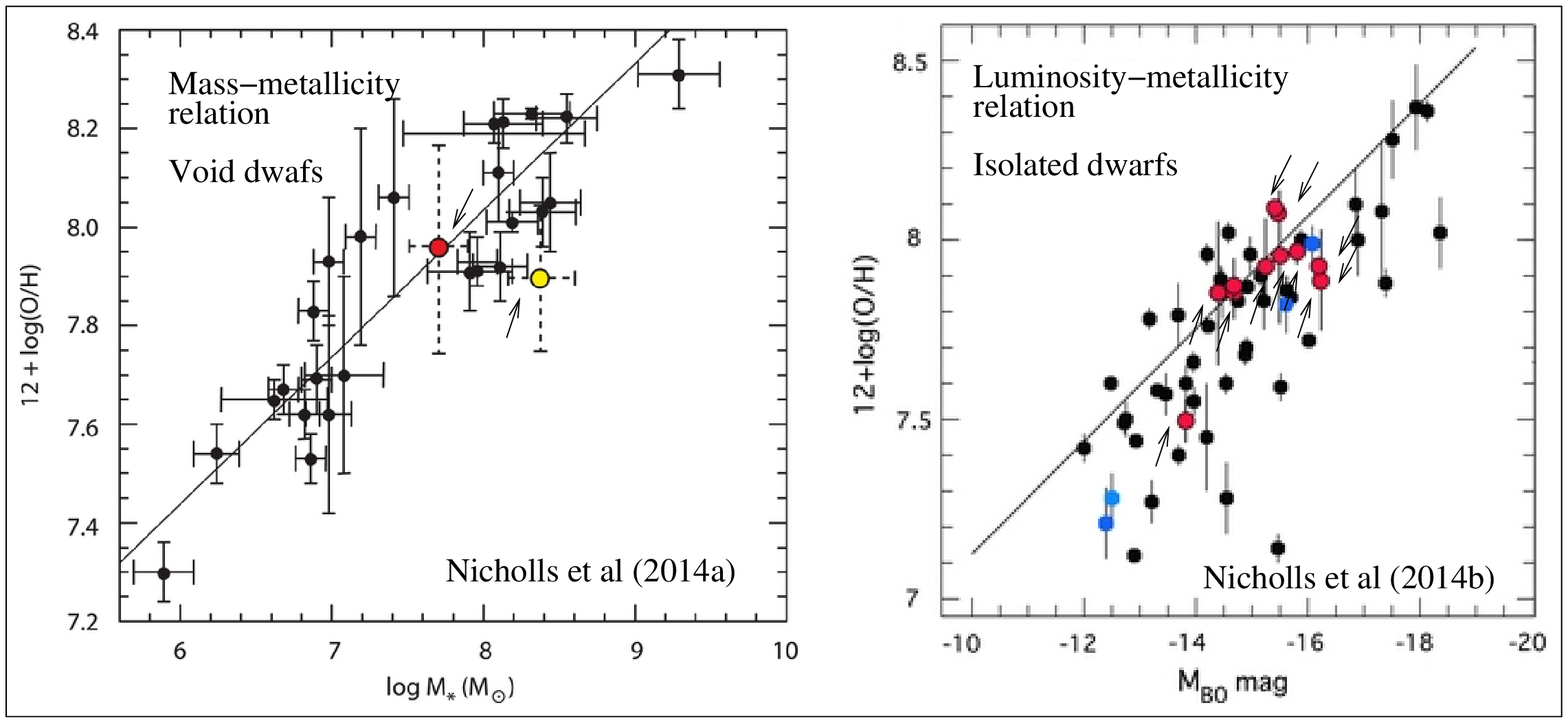} 
 \includegraphics[width=5.0in]{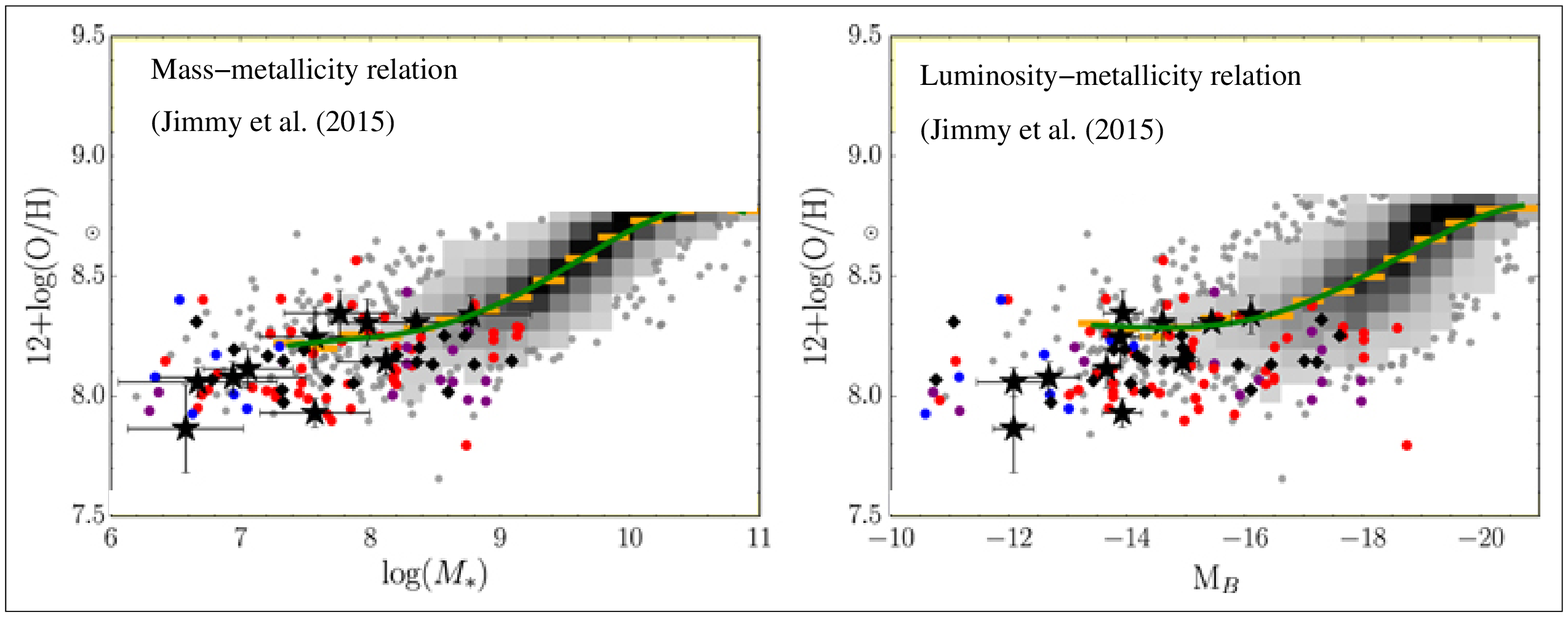} 
 \caption{{\it Top:} Luminosity and mass-metallicity relations for the void and isolated dIrr galaxies from SIGRID (yellow and red circles, also marked with arrows), as compared to the relations for other dwarf galaxies, from Nicholls et al. (2014a, b). {\it Bottom}: The MZR and LMR based on optical integral-unit field VIMOS$@$VLT data, to account for the low mass end of the relations. The plots show points from different sources to emphasize the similarity of the relations. IFU data points are represented by the star symbols (Jimmy et al. 2015).
}
   \label{f7ul}
\end{center}
\end{figure}

An alternative way to investigate the universality of the MZR is by inspecting its validation in a variety of contexts. In Figure~\ref{f7ul} the MZR of dwarfs is explored in three ways, and quickly described below.

{\it i})
Based on the isolated gas-rich irregular dwarf galaxies from SIGRID (The Small Isolated
Gas Rich Irregular Dwarf), Nicholls et al. (2014a, 2014b) found out that void and isolated dwarf irregular galaxies follow the same MZR as normal dwarfs of the same range of masses. See the top panel of Figure~\ref{f7ul}. 

{\it ii})
 Very simple (one parameter) chemical evolution models with multiple star forming bursts,  for Draco, Leo~I, IC~1613 and Fornax can explain fairly well the empirical stellar MZR obtained by Leaman (2012) and Kirby et al. (2013), on which dIrr and dSph galaxies follow the same relation (Hartwick 2015).
       
{\it iii})
 The dwarf galaxies MZR is investigated by using VIMOS-IFU on the VLT and ALFAFA (Arecibo Legacy Fast ALFA, a blind H~{\sc i} survey of the local universe) to explore the role of H~{\sc i} in the MZR. This gas-phase MZR is found out to continue down to stellar masses as low as 10$^{6.6}$~M$_{\odot}$ (Jimmy et al. 2015).  See the bottom panel in Figure~\ref{f7ul}.

From the discussion in this section, the LZR and the MZR reveal that some physical mechanism must be driving a correlation between stars, metals, and gas flows across a vast range of scales, although, it is still uncertain the nature of the mechanism. Let us then emphasize that both dIrr and dSph galaxies are consistent with the global MZR of star forming galaxies, likely because chemical evolution is a function of stellar mass and its correlation with the total (baryonic and non-baryonic) mass of the galaxy.

\section{Remarkable conclusions to keep in mind}

-- PNe are present in star-forming and in quiescent galaxies, so they provide homogeneous metallicities for all the Local Group dwarf galaxies.

-- The third dredge-up could imply that oxygen from PNe is not always a good metallicity indicator. Fortunately the 3du is rare, and only could occur for galaxies with 12+log(O/H)$\leq$7.7. In these cases (O/H)$_{PNe}$ does not allow galaxy evolution studies.

-- Models with PNe and H~{\sc ii} regions as constraints also agree with the multiple SF bursts and winds that account for the evolution of dIrr and dSph galaxies.

-- Metallicity distributions based on the gas-phase abundance are available for a number of LG dSph, dIrr, early- and late-type spiral galaxies. However, so far it is not clear if the time (PNe and 
H~{\sc ii} regions) evolution of the O/H gradients represent “slight” steepening or flattening. 

-- The H~{\sc ii} regions, planetary nebulae, and the old stars all suggest that the mass-metallicity relation is universal. The relation holds from ~10$^{6.6}$~M$_{\odot}$ to ~10$^{11}$~M$_{\odot}$, by spanning the mass range of the LG dwarf irregulars, dwarf spheroidals as well as the SDSS star-forming galaxies. 

\bigskip
\bigskip
\bigskip
I would like to acknowledge S. Stierwalt \& K. McQuinn for so enthusiastically organize this symposium, which had many great and enlightening discussions. I also thank the IAU and CNPq (304184/2016-0 and 451660/2018-8) grants.

{}

\begin{discussion}

\discuss{Thiago S. Gon\c calves}
{Can the same metallicity proxies, PNe and H~{\sc ii} regions, be used to derive the MZR at high redshifts?}

\discuss{Denise R. Gon\c calves}
{In fact the PNe oxygen abundances can be measured at most up to 4~Mpc. With H~{\sc ii} regions, which are usually much brighter than the PNe, the situation is different. However, the MZR of PNe is that robust, exactly because it is obtained with the same Z indicator. And, as it is discussed in  
this review, deep enough spectra are needed, which prevents this exercise at high redshifts.}
  
\discuss{Lee?}{You brought up the important point that oxygen is not always a good proxy for metallicity because of the evolution of progenitor star used as an abundance probe. However, because of chemical evolution of galaxies, oxygen is not uniquely linked to Fe either. How does this affect the mass-metallicity relation?}

\discuss{Denise R. Gon\c calves}{The fact that sometimes Fe is used as a proxy for metallicity and sometimes O is, will notably be included in the scatter of the mass-metallicity relation. PNe are relevant to derive the MZR exactly because they allow to avoid this conversion! O/H is safe, provided that for galaxies with 12+log(O/H)$\leq$7.7 one takes into account the trends of the $\alpha$-elements like S and Ar wrt O, thus being sure that this element was not produced by the progenitor stars of the PNe in galaxy considered.}

\end{discussion}

\end{document}